# On The Strength of Tribo- Emission in Sliding of Diamond on Single Crystal Silicon

Hisham A. Abdel-Aal*

Laboratoire de Mécanique et Procédé de Fabrication (LMPF, EA4106)
Arts et Metier ParisTech, LMPF, Rue Saint Dominique BP 508, 51006, Chalons-en-Champagne - France.

*corresponding author
hisham.abdel-aal@chalons.ensam.fr

**Abstract**

Triboemission is a phenomenon associated with the sliding of variety of materials. The phenomenon is thought to be related to wear of diamond tools used in precision machining of semiconductors. As such, the physics of emission has recently acquired importance. Many researchers studied emission during scratching of solid surfaces. They observed that the intensity of tribo-induced emission of the electrons, ions, and photons decrease in the order: insulator>semiconductor> conductor. Many experiments conducted to compare the emission of negatively-charged particles in case of the semiconductors Si with that from selected insulators have reported a clear decreasing trend of the tribo-emission intensity as contact progressed over the same wear track for diamond-on-Si. Despite that all of these experiments were performed in vacuum, the origin of the weak signals and the decrease of signal strength in the case of Si centred on the presence of dielectric silicon oxide films formed in air during surface preparation or because of the samples being of mono-crystalline nature. This paper offers an alternative explanation to the behaviour of tribo-emitted particles based on the pressure induced semi-conductor-to-metallic phase transformation that takes place in Si during sliding. It is shown that due to repeated sliding, the wear tracks experience a semiconductor-to- metal transformation that renders the bulk of material immediately under the diamond slider conducting.

**Biographical notes**

Hisham A. Abdel-Aal obtained his Doctoral Degree from the University of North Carolina in Charlotte in Mechanical Engineering. He is currently an Invited Research Professor at Arts et Métier ParisTech. His research interests fall within the Tribology and Thermodynamics of solids.

**Introduction**

Triboemission is a phenomenon associated with the sliding of variety of materials including: metals, non-metal elements, ceramics, glasses, and rocks, in various environments, e.g., in ambient air, in various gases under atmospheric pressure as well as in vacuum and in lubricating oil. (Nakayama, et. al, 1990, 1995, 1997; Miura and Nakayama, 1994, 1999, 2000, Dickinson, et., al 1997, Chapman and Walton, 1983, Nakayama and H. Hashimoto, 1995, Molina et., al, 2001). The phenomenon is thought to be related to wear of diamond tools used in precision machining of semiconductors. As such, the physics of emission has recently acquired importance. Several researchers investigated triboemission from sliding semi-conductor surfaces. Dickenson et al., (Dickenson et al., 1997, 1987), measured electron emission outputs from bending fracture of single crystal Si and atomic and molecular emission from Ge fractured surfaces. Kaalund and Haneman (Kaalund and Haneman 1998) studied particle emission during the cleavage-by-bending of Si and Ge cantilevers in vacuum. They observed burst type electron emission starting at the onset of cleavage. The duration of the emission bursts ranged from tens of microseconds to around 1.8 milliseconds. Moreover, burst signals were independent of dopant concentrations, high vacuum levels and, of temperatures. Nakayama and Fujimoto (Nakayama and Fujimoto 2004) monitored the distribution of electron emission intensity during scratching the surface of the n-type semiconductor Si (100). They also observed monotonic decay of the intensity of electron emission with repeated sliding. These authors attributed the decrease of electron emission to the mono-crystalline nature of the silicon samples used in the experiments. Their reasoning stemmed from observation of similar emission behaviour in the sliding of diamond against crystalline $Al_2O_3$ and MgO. They, however, noted that the energy of the electrons emitted from semi conductive silicon is lower than that emitted from sintered $Al_2O_3$ by an order of magnitude due to the lower electrical resistivity of silicon than that of $Al_2O_3$. Molina et al., (Molina et al., 2004, 2002) compared triboemission from semiconductors, Si and Ge, with that of selected insulators. In agreement with Nakayama and Fujimoto they reported that for diamond-on-Si and diamond-on-Ge tribo-pairs emission intensity rapidly decreased as the contact progressed over the same wear track. They also reported as absence of triboemission from the semi-conductors Si and Ge after the contact have ceased (in comparison to significant post contact emission from insulators). They hypothesized that the absence of surface charge on semi-conductors in vacuum relates to the observed absence of emission.

The origin of the observed triboemission behaviour is still unclear (Molina et al 2007). There are several proposed explanations with none widely accepted. These are variations on the reduction of surface charge theme. Therefore the crux of such proposals is an explanation of the origins of change in the work function of the solid while sliding. Dickinson et al (1983) proposed that local straining would reduce the work function and thereby leads to charge separation. (Molina et al 2007), upon observing plastically extruded material along the wear tracks in the sliding of Si and Ge, proposed that plastic deformation is the cause of the hypothesized reduction in the work function.

The extensive experiments of Nakayama and Fujimoto (2004) suggested that the strength of triboemission follows the hierarchy of electrical resistivity of the sliding solid. That is the intensity of tribo-induced emission of the electrons, ions, and photons decrease in the order: insulator>semiconductor> conductor. Plastic deformation in semi-conductors, and ceramics, meanwhile generally takes place due to Pressure Induced Metallization (PIM). Thus, for these materials, which generally respond to mechanical loading and indentation in a brittle manner, deform plastically if the local pressure (stress) reaches a critical value. Upon reaching this critical state of stress, which is approximately equal to the mechanical hardness, the affected volume of the material exhibits a metallic response that is marked by a local reduction in the electrical resistivity. That is, the volume of the material that experiences a mechanical load equal to the hardness, originally covalent, becomes conductive. Such a phenomenon may offer an alternative view of the origin of tribo-emission behavior in the sliding of semi-conductors and ceramics. That is, combining the hierarchy of tribo-emission strength and the PIM change in resistivity may explain the fading of tribo-emission observed in sliding of semi-conductors. This work, therefore, investigates such a proposal. Here we explain the experimentally observed rapid decay of emission during sliding based on pressure-induced metallization.

The organization of the paper is as follows: the first part describes the high-pressure behavior of Si and provides a summary of electrical transport properties of the High Pressure Phase Si-II ($\beta$–tin). In the second part of the paper, we use the available data to predict and explain the observed tribo-emission behavior of silicon in scratching.

## 2. Behavior of silicon in scratching

High Pressure Phase Transformation (HPPT) phenomenon in semiconducting and ceramic materials has been investigated for over 50 years. It is settled that under very high pressures, comparable to their hardness, many covalent materials transform to a metallic state (Bridgman, 1948). Indeed, X-ray diffraction studies showed a series of phase transformations that materials undergo depending on the material and deforming conditions. Among these, the most relevant to the subject of the current study is the so-called covalent-to-metallic phase transformation (Jamieson, 1963, Hu, 1984).

Mono-crystalline semiconductors are generally considered fragile in that they exhibit brittle response under normal loading conditions. However, when the mechanical loading reaches a critical threshold, they undergo structural changes that lead to ductility. Discovery of this behavior lead to intensive investigation of the changes induced by high pressures (George, 1999, Drickamer 1970, Hall et., al, 1963, Hu et. al, 1986, Kasper and Richards 1963). Results have shown that under hydrostatic loading conditions, at room temperature, both Ge and Si, once the loading pressure reaches the mechanical hardness, change from semi-conducting to metallic takes place. That is, the so called Si-I and Ge-I phases, cubic lattice, if subjected to a pressure of 8.5-16 GPa for the former and 10-11 GPa for the later, which is equivalent to the hardness, will transform into the $\beta$-tin metallic structures Si-II and Ge-II. Germanium and Silicon undergo similar transformations, thus, in what follows we are going to focus on the behavior of Si without loss of generality.

### 2.1 pressure induced phases of Si

At atmospheric pressure, silicon has a cubic diamond structure (space group Fd3m) up to the melting temperature. Conventionally, this phase is labeled Si-I. At elevated pressures, 11 other crystalline phases of Si have been identified by calculation and from pressure cell experiments (Kasper and Richards 1964). Of interest for the purpose of this work is the phases referred to as Si-II (f3-tin structure, space group I41 /amd) (Kobliska and Solin, 1973, Gogotsi et., al 2000) which exhibits metallic (conducting) behavior (Gogotsi et., al, 2000, Kovalchenko2002). The transition from Si-I to Si-II occurs in the pressure range of 8.5 GPa < P <16 GPa. This transition is not reversible in the sense that Si-II transforms to different metastable phases depending on pressure release conditions. The Si-I to Si-II transformation is also followed by ~20% densification of the material (Domnich and, Gogotsi, 2002).

Figure 1 shows a schematic illustration of the phase transformations cycle in silicon under hydrostatic contact loading.

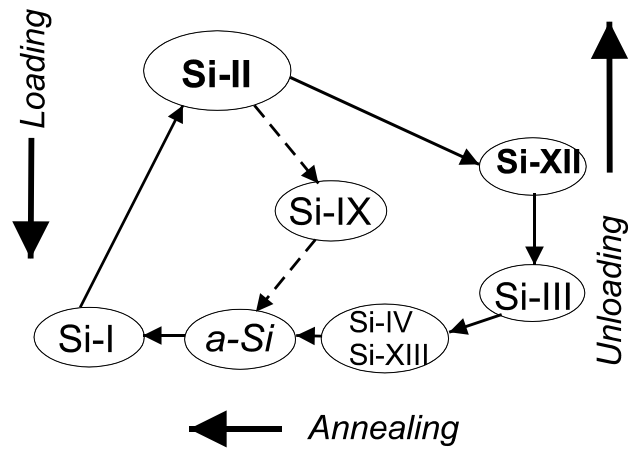

*Figure 1* schematic illustration of the phase transformations cycle in silicon under hydrostatic contact loading

### 2.2 Scratching and wear behavior

Gogotsi et al. (2001) conducted a comprehensive study of the response of Silicon during scratching. The authors used sharp (Vickers) and blunt (Rockwell) diamond indenters for scratching. Post-scratching phase and stress analysis was performed by means of Raman micro- spectroscopy, and groove morphology (characterized by scanning electron microscopy), atomic force microscopy and, optical profilometry. All post mortem analysis methods confirmed that Silicon response to scratching in the ductile regime proceds along the following paths: Highly localized stresses underneath the tool lead to the formation of the metallic phase Si-II (β-tin). This phase deforms by plastic flow and subsequently transforms into a-Si or a mixture of Si-III and Si-XII in the wake of the scratching tool depending on speed of loading and unloading. These reverse phase transformations are accompanied by a ~10% volume increase.

Using Raman spectroscopy as a primary characterization tool, Kovalchenko et al. (Kovalchenko, 2002) investigated the possibility of the metallization of Si during tribological situations. Standard ball-bearing silicon nitride balls (9.55 mm in diameter) were rubbed against single-crystal silicon wafers at linear velocities of 0.4 to 30 mm/sec. Such low values of linear velocities were selected in order to diminish the extent of frictional heating and thus to suppress possible tribo-chemical reactions. Variable test duration allowed investigation of the wear tracks on silicon after 1 to 300 runs under loads of 1 to 20 N. Applied load and the number of runs appeared to be the important parameters that determine the mechanism of Si wear during tribological experiments.

Scan Electron Microscopy (SEM) images of selected areas within the wear track on silicon subjected to an interrupted 5-run friction test (applied load 1N) clearly revealed evidence of some plastic deformation and seizure. The Raman spectra taken from the corresponding areas have pronounced features associated with amorphous Si and the crystalline Si-III and Si-XII phases the presence of which in the wear tracks suggests partial pressure-induced metallization of Si during the open-air friction tests similar to other types of contact loading. In particular, the appearance of the plastic extrusions in the wear tracks indicates to the formation of a ductile metallic Si-II phase and its plastic flow under a moving Si3N4 ball. Figure 2 presents a schematic illustration of phase configuration during scratching of silicon with a sharp diamond indenter.

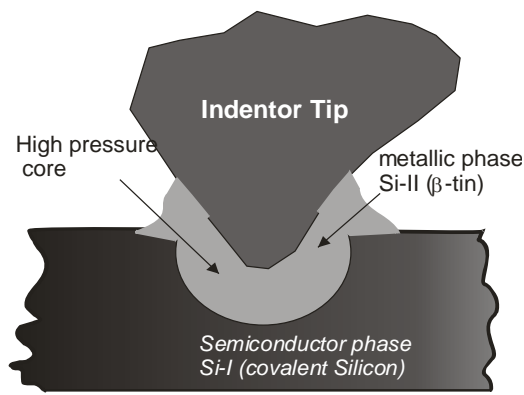

Figure 2  phase configuration during scratching of silicon with a sharp diamond indenter.

Note that the PIM transformation, i.e the formation of the metallic phase Si-II, is confined to the material volume directly located under the indenter. That is the transformation takes place under the indenter within the high-pressure core.

## 2.3 Pressure induced changes in electric transport properties

Silicon in its semi-conducting (covalent) state is a poor electrical conductor (its' electrical resistivity is around 1E04 $\Omega$.m). When a sharp indenter scratches Semiconductor silicon, a high-pressure zone characterized that experiences high-localized stresses will form under the tip of the stylus. The electrical resistivity within this zone is typically drops drastically when the local pressure falls within the critical threshold 8.5 GPa < P < 11 Gpa. The drop in the electrical resistivity indicates a rise in the ability of the high-pressure core for electrical conduction. As such, when the drop in resistivity takes place it is commonly interpreted as an indication that the semiconductor, covalent, phase Si-I has transformed to a conducting metallic phase (the so called Si-II $\beta$-tin). That is the surface layer, originally semi-conducting, has transformed into a metallic layer. A typical trace depicting the change in resistance while indentation, recorded in situ (Abdel-Aal, Patten and Dong, 2005), is shown as figure 3.
Combining pressure and heat may also induce a transformation into a metallic state (although the mechanism responsible for the transformation in this case is quite different from that inducing the change due to hydrostatic pressure (Abdel-Aal, Patten and Dong 2005, Abdel-Aal et al. 2006). Figure 4 is a summary plot of the change in electrical resistivity of the metallic phase (Abdel-Aal et al. 2006) with temperature. The figure indicates that when the drop in resistivity increase with temperature elevation (when the critical pressure is kept constant). The highest value being that when no temperature rise is involved. Note also that the value of the resistivity, regardless of the value of temperature rise, is considerably less than that of the covalent Si. The results presented in the figure, generally support the findings of other researchers who investigated the change in resistivity as a function of hydrostatic pressure only. Clarke et al (1988), reported qualitatively that the electrical conductivity of Si increases by two orders of magnitudes due to metallization. This implies that a good estimate for the resistivity, which is the reciprocal of the conductivity, will be a drop by one-half of that of the semi-conducting phase. Gilev and Trubachev (1999), measured the effect of pressure on the resistivity of Si in shock compression. They reported that the resistivity falls in the interval *2 E-05 $\Omega$cm < $\rho$ < 2.5 E-05 $\Omega$cm*. Bundy and Casper (1970) reported a value of $\rho$ = 6E-06 $\Omega$cm under a static pressure of 15 Gpa which, again, is smaller than that of the control sample used in their experiments and smaller than the nominal value for the covalent Si ($\rho$ = 1E04 $\Omega$.m) which is commonly quoted in literature. In all we can infer from the data that the effect of silicon metallization, i.e., the transformation to the Si-II phase due to pressure, is a drop in the electrical resistivity of the new phase compared to the covalent phase Si-I.

## 3. Implications for tribo-emission

The energy of emission depends on the electric field generated by tribo-charging (surface potential). This, on the other hand, is a function of two parameters, the electrical resistivity, and the dielectric constant (which relates to the work function of the solid) with the resistivity being the dominant factor.

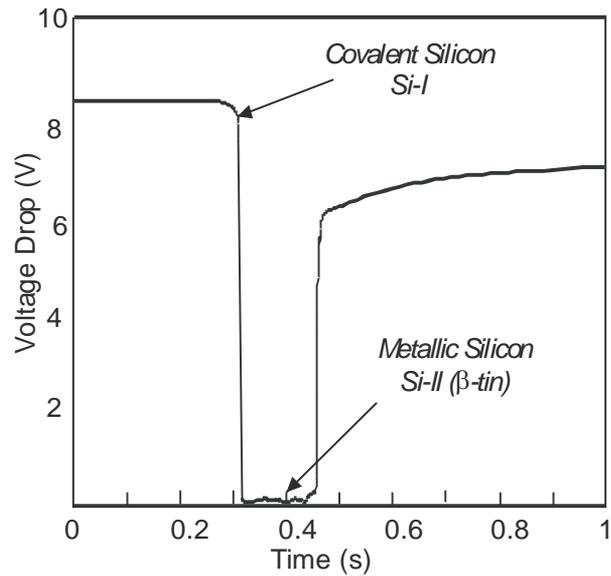

Figure 3 Typical voltage trace from a scratching experiment in which a Single crustal Diamond indenter is used to scratch a Silicon wafer. Experiental conditions are: scratching load 0.03 N, Speed 0.1 mm/s.

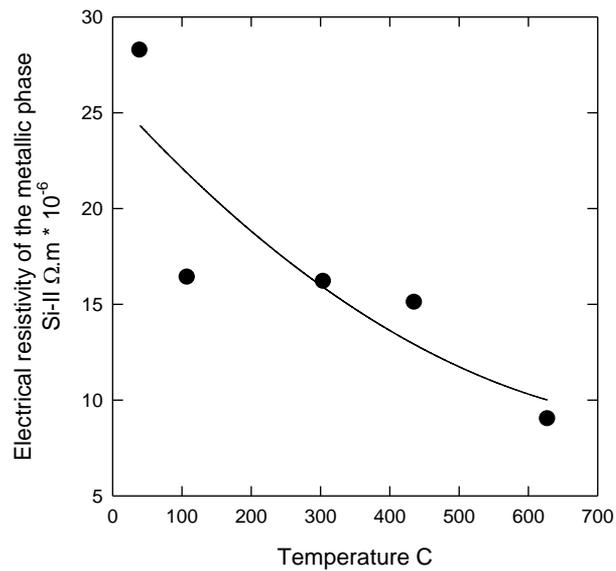

Figure 4: Variation in the electrical resistivity of the metallic phase Si-II with temperature

The surface potential, V, is calculated from (Nakayama 1999):

$$V = \frac{k}{\varepsilon} \exp\left(\frac{-t}{\rho \varepsilon}\right) \quad (1)$$

Where, $k$ is a constant, $\varepsilon$ is the dielectric constant, $t$ is the emission time and $\rho$ is the electrical resistivity.

Equation (1) indicates that when the resistivity of the material decreases (which is synonymous with the increase in the electrical conductivity since the conductivity is the reciprocal of the resistivity) then the surface potential V would weaken. Moreover for a metallic material the dielectric constant is zero and such a value would render the surface potential of zero strength whence no emission.

The strength of emission is directly proportional to the strength of the surface potential generated by tribo-charging. This allows using equation one to compare the strength of emission emanating from each pase (semi-conducting and metallic). To this end, we write equation one for the semiconducting phase and then divide that by the equation written for the metallic phase. This yields a ratio that compares, qualitatively at least, the strength of emission emanating from each of the phases at any given moment within a wear (or scratching cycle). As such, first we assume that the constant k is phase independent (i.e., its' value is the same for both phases). We further assume that the PIM does not affect the dielectric constant, $\varepsilon$. Under these assumptions, the ratio between the strength of particle emission from the semi-conducting phase (Si-I) to that from the metallic phase (Si-II) is expressed as:

$$\frac{V_{sc}}{V_m} = \exp\left[-\frac{t}{\varepsilon}\left(\frac{1}{\rho_{sc}} - \frac{1}{\rho_m}\right)\right] \quad (2)$$

Where the subscripts: $sc$ and $m$ denote the semi-conducting phase (Si-I) and the metallic phase (Si-II) of silicon respectively.

Equation (2) was evaluated using the electrical resistivity of the semi-conducting silicon phase Si-I, 1E04 (Harper 2001), the resistivity of the metallic phase $\rho_m$ for three temperatures (Abdel-aal et., al 2006) that are typical of the temperature rise exhibited by the silicon samples in slow sliding experiments; and the dielectric constant $\varepsilon=12$ (Nakayama 1999). Figure (5) depicts the change in the ratio of surface potential (and thereby the strength of emission) $V_{sc}/V_m$ computed from equation two. Table 1 presents a summary of the values used in computations.

Table 1 Summary of properties used in computing results presented in figure (5).

| Phase | | Resistivity $\Omega$.m |
|---|---|---|
| Si-I (covalent) | | 1E-04 |
| Si-II (metallic) | **Temp (C)** | |
| | 5 | 5.22E-05 |
| | 10 | 4.19E-05 |
| | 15 | 3.67E-05 |

The figure shows that for all temperatures the surface potential of the metallic phase is of comparable magnitude to that of the semi-conducting phase only for short emission times (typical of the first few milliseconds at the initiation of sliding *$10^{-8} < t < 10^{-4}s$*). For emission times greater than around $10^{-3}$ seconds, the surface potential for the metallic exhibits a sharp drop. This implies that if PIM takes place the emitted particles will not be of detectable strength beyond the $10^{-3}$ second threshold from the beginning of any wear cycle (i.e., one complete revolution on the wear track). It is interesting to remark that such a time interval $0 < t < 10^{-3}$ was the observed window of detectable emission burst observed in several experiments. The experimental observations of Nakyama (Nakayama 1999) provide additional supporting evidence for the effect of PIM on tribo-emission. These experiments entailed the use of sharp diamond indenter, which is likely to induce a stress field of a magnitude greater than the critical value necessary for PIM. In fact, their experimental conditions into the Hertz contact equation, and the Johnson solution for a conical indenter respectively (Johnson 1987), the contact pressures are found to fall between the critical pressures that favor PIM ($8.5 < P < 11$ GPa). Molina (2004) also observed plastically extruded silicon along the wear track, which is an additional indication of a PIM taking place.

The combination of: a contact pressure that falls in the critical range and the observation of plastically extruded material, strongly suggests that PIM is the mechanism responsible for the emission decay observed in the experiments. This is because the intensity of the emitted particles is a function of the surface potential. This, in turn, is a strong direct function of the electrical resistivity of the solid. Moreover, undetectable (or weak) emission is a feature of a conductor (metallic material) possessing low electrical resistivity. So for a semi-conductor to start emitting particles of detectable strength, then shortly thereafter this emission becomes undetectable, a drop in electrical resistivity has to take place. Such a drop is only possible, if the contact stress favors a PIPT that leads to metallization of the material volume directly affected by the localized pressure.

The extensive experiments of Nakayama and Fujimoto (Nakayama and Fujimoto 2004) resulted in a suggested hierarchy of triboemission intensity order of magnitude from different materials. The authors suggested that the energy of tribo-emitted electrons in a normal atmosphere should follow the hierarchy insulator>semiconductor> conductor. The effect of PIM is to change the electrical resistivity from a value typical of a semi-conductor to a value typical of a metallic material. The observed behavior of emission, moreover, follows the hierarchy suggested by Nakayama and Fujimoto. Using this observation in conjunction to the computed values of the ratio of emission strength, we can propose the following to explain for the observed trend of emission intensity.

Upon contact between slider and Si-sample, a stress field will develop. Initially, the magnitude of the stress will be less than the critical value necessary to induce metallization. This allows emission to take place. The pressure in the high pressure core, the MAZ, will evolve with time. As such, during the period necessary for evolution of pressure, parts of the MAZ will experience a stress magnitude less than the critical value. This allows detectable particle emission to take place from such zones. Once the entire stress field within the MAZ reaches the critical threshold weak, undetectable, emission will take place. However, the source of this emission is not necessarily the material volume within the MAZ.

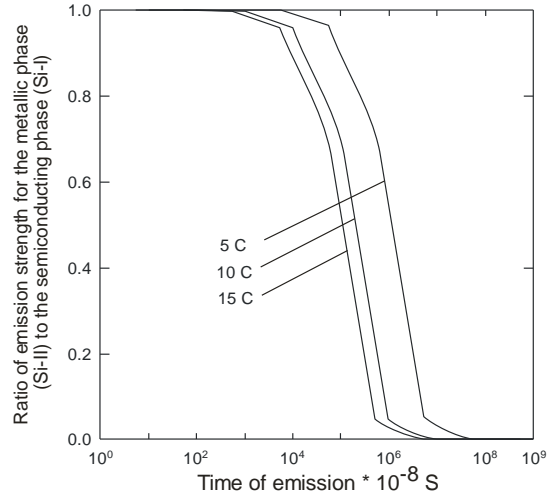

Figure 5  The ratio of emission strength of the metallic phase Si-II to that of the semi-conducting phase Si-I ,at three different temperatures, evaluated for different emission times.

Due to repeated sliding, the diamond stylus will expose new material in the wear track. The magnitude of the pressure acting on the newly exposed material is not necessarily that which is needed for PIM. Whence, the newly exposed material volume will emit detectable emission. When a material volume within the MAZ reaches the critical stress value, metallization of this particular volume will take place. The transformed material will flow sideways, by plastic extrusion, in the wear track. This will relieve the stress acting on the material flowing sideways and will give rise to a back transformation that leads to the formation of the amorphous phase a-Si (refer to figure 1). The a-Si phase is not of a metallic nature. As such, when the extruded material volume rubs against the indenter, or the diamond slider, it will emit particles.

There is a need for more experimental observations of emission at higher speeds and variable pressures. Also, of interest, are experiments where sliding continues for longer distances (i.e more revolutions on the same wear track). This is because of the dominating effect of PIM in the early stages of sliding (n< 300 revolutions (Kovalchenko 2002)-compare to the 20 revolutions used by Nakayama).

**CONCLUSIONS**

This paper suggests that the origin of the behavior of the Tribo-induced emission in Silicon observed in scratching experiments is linked to the pressure induced semi-conductor-to-metal phase transformation. It has been shown that due to the

*Title*

drop of the electrical resistivity of the metallic phase Si-II, compared to that of the covalent phase Si-I, a significant drop in the surface potential will take place. This will result in a very weak particle emission from the scratched surfaces.

**Nomenclature**
HPPT    High Pressure Phase Transformation
MAZ     Mechanically Affected Zone
PIM     Pressure Induced Metallization
SC      semi conductor

*Title*